\documentclass[aspectratio=169]{arxiv}
\usepackage{moresize}
\usepackage{setspace}
\makeatletter
\renewcommand\huge{\@setfontsize\huge{30pt}{30}}
\renewcommand\HUGE{\@setfontsize\HUGE{80pt}{40}}
\makeatother
\PassOptionsToPackage{svgnames,table,dvipsnames}{xcolor}
\usepackage{xcolor}
\PassOptionsToPackage{breakable,xparse}{tcolorbox}
\definecolor{shadecolor}{gray}{.95}
\definecolor{darkred}{rgb}{0.3, 0.0, 0.0}
\definecolor{darkgreen}{rgb}{0.0, 0.3, 0.1}
\definecolor{darkblue}{rgb}{0.0, 0.1, 0.3}
\definecolor{darkorange}{rgb}{1.0, 0.55, 0.0}
\definecolor{sienna}{rgb}{0.53, 0.18, 0.09}
\definecolor{ForestGreen}{rgb}{0.13, 0.55, 0.13}
\usepackage{newverbs}
\newverbcommand{\rverb}{\color{darkred}}{}
\newverbcommand{\gverb}{\color{darkgreen}}{}
\newverbcommand{\bverb}{\color{darkblue}}{}

\makeatletter\@ifclassloaded{beamer}{}{\@ifclassloaded{arxiv}{}{\usepackage[letterpaper, top=0.5in, bottom=0.5in, left=0.5in, right=3in]{geometry}}}\makeatother

\usepackage{tikz}

\usetikzlibrary{positioning, arrows.meta}
\author{%
Adolfo De Unánue\orcid{0000-0003-3499-0965} \\
Centro de Ciencia de Datos e Inteligencia Artificial \\
Escuela de Gobierno y Transformación Pública \\
Tecnológico de Monterrey \\
\email{unanue@tec.mx}
\And
Fernanda Sobrino\orcid{0000-0001-8901-6022} \\
Centro de Ciencia de Datos e Inteligencia Artificial \\
Escuela de Gobierno y Transformación Pública \\
Tecnológico de Monterrey \\
\email{fersobrinno@tec.mx}}
\shorttitle{ML as Materialist Practice}
\shortauthor{De Unánue and Sobrino}
\date{2026-05-13}
\title{Machine Learning as Performative Materialist Practice\\\medskip
\large Thirteen Theses on the Epistemology, Methodology, and Politics of Applied ML}
\usepackage{calc}
\newlength{\cslhangindent}
\newlength{\csllabelsep}
\newlength{\csllabelwidth}
\newenvironment{cslbibliography}[2] 
 {
  \setlength{\parindent}{0pt}
  \ifodd #1
  \let\oldpar\par
  \def\par{\hangindent=\cslhangindent\oldpar}
  \fi
  \setlength{\parskip}{\parskip +  #2\baselineskip}
 }%
 {}

\newcommand{\cslbibitem}[2]
  {\leavevmode\vadjust pre{\hypertarget{citeproc_bib_item_#1}{}}#2}
\makeatletter
\newcommand{\cslcitation}[2]
 {\protect\hyper@linkstart{cite}{citeproc_bib_item_#1}#2\hyper@linkend}
\makeatother
\begin{document}

\maketitle
\begin{abstract}
Machine learning practice in \emph{institutional decision-support contexts}---government, public policy, public health, criminal justice, resource allocation---rests on a set of largely unexamined epistemological commitments inherited from classical statistics and computer science: that models represent stable regularities, that validation can be context-free, that performance metrics are politically neutral, and that feature importance reveals system structure. This paper challenges these commitments through a unified framework of \emph{performative materialist ML}, articulated as thirteen theses. Drawing on Pickering's cybernetic ontology, the performativity literature from economic sociology (Callon, MacKenzie), Simon's bounded rationality, the formalization of performative prediction (\cslcitation{27}{Perdomo et al. 2020}), and fifteen years of applied ML experience in government and public policy, we argue that: (1) ML models are best understood not as truth-seeking representations but as temporally situated \emph{compressions} that function as instruments of intervention; (2) the full data product is a complex adaptive system that coevolves with its target and navigates a multi-objective space no single algorithm can optimize; (3) validity is fundamentally performative, measured by effects in the world rather than formal properties of the model; (4) the choices embedded in objective functions, fairness criteria, and resource thresholds are political decisions belonging to stakeholders, not technicians. We show how these theses unify several practical prescriptions---temporal cross-validation, precision and recall at k, pipeline-aware fairness auditing, satisficing over optimizing---as consequences of a coherent materialist epistemology rather than isolated best practices.
\end{abstract}

\section{Introduction}
\label{sec:orgb0b4799}

The practice of machine learning has developed a remarkably robust set of tools and an equally impoverished philosophical self-understanding. Practitioners routinely deploy models that reshape the systems they predict, evaluate them with metrics disconnected from the decisions they inform, interpret feature importance as if it revealed causal structure, treat validation protocols as interchangeable regardless of temporal context, and optimize single loss functions as if the problem were unidimensional. These are not merely technical oversights; they reflect deep epistemological commitments that remain largely invisible to the community that holds them. To make this concrete: when a prosecutor's office deploys an ML score to prioritize cases (\cslcitation{40}{Sobrino et al. 2026}), it is not \emph{observing} a static distribution of case viability---it is \emph{modifying} that distribution by intervening on which cases get attention, which witnesses get interviewed, which evidence gets analyzed. The model participates in the system it describes, and that participation is precisely what the standard epistemology cannot see.

This paper proposes a unified framework---what we call \emph{performative materialist ML}---that makes these commitments explicit, subjects them to critique, and replaces them with a coherent alternative. The framework is articulated as thirteen theses, organized around four dimensions: epistemology (what models know), ontology (what models are), methodology (how models should be validated), and politics (whose interests models serve).

The intellectual roots of this framework are diverse. From Andrew Pickering's account of British cybernetics, we draw the distinction between representational and performative ontologies (\cslcitation{28}{Pickering 2010}). From the economic sociology of Callon (\cslcitation{10}{Callon 1998}, \cslcitation{11}{2007}) and MacKenzie (\cslcitation{23}{MacKenzie 2006}), we draw the insight that models actively constitute the systems in which they are deployed. From Simon's (\cslcitation{38}{Simon 1956}) theory of bounded rationality, we draw the concept of satisficing as the appropriate decision strategy in complex environments. From the formalization of performative prediction by Perdomo et al. (\cslcitation{27}{Perdomo et al. 2020}), we draw the mathematical demonstration that prediction and intervention are entangled. And from fifteen years of applied ML in government contexts---including the DSSG program, the Triage pipeline, and the Aequitas fairness audit toolkit (\cslcitation{35}{Saleiro et al. 2018})---we draw practical lessons that are better understood as consequences of a materialist epistemology than as isolated best practices.

This paper makes three contributions. First, it provides a philosophical foundation for a set of widely known methodological prescriptions: the primacy of temporal validation, the irrelevance of global metrics, the inadequacy of feature importance for decision support, the necessity of satisficing over optimizing, the multi-objective nature of all real ML problems, and the inseparability of technical and political choices. Second, it connects the applied ML literature to intellectual traditions---cybernetics, economic sociology, historical materialism, bounded rationality---that have remained almost entirely outside the field's awareness. Third, it proposes a framework for evaluating ML systems that is explicitly performative and political.
\section{Related Work}
\label{sec:orgf8c6038}

\subsection{Breiman's Two Cultures and Beyond}
\label{sec:orgfbd4ce6}

Breiman's (\cslcitation{9}{Breiman 2001}) distinction between the ``data modeling culture'' and the ``algorithmic modeling culture'' identified a fundamental divide in statistical practice. The algorithmic culture is the intellectual ancestor of modern ML. However, Breiman's framework remains epistemologically conservative: it treats prediction as the end goal rather than as instrumental to action, and assumes that predictive accuracy can be evaluated through cross-validation on shuffled data. We propose a ``third culture'' that goes beyond prediction to intervention and beyond accuracy to performative validity.
\subsection{The DSSG/Triage Tradition}
\label{sec:org03248dc}

The DSSG program (\cslcitation{31}{Rodolfa et al. 2019}) developed a distinctive approach to applied ML for public policy that implicitly embodies several of our theses. Ackermann et al. (\cslcitation{1}{Ackermann et al. 2018}) codified the program's operational discipline as an explicit deploy-monitor-retrain framework, directly anticipating the coevolutionary structure of Thesis II. The program's empirical record spans criminal justice (\cslcitation{18}{Helsby et al. 2018}; \cslcitation{12}{Carton et al. 2016}; \cslcitation{6}{Bauman et al. 2018}), public health (\cslcitation{29}{Potash et al. 2015}; \cslcitation{22}{Kumar et al. 2020}), social services and government response (\cslcitation{36}{Sankaran et al. 2017}; \cslcitation{15}{Gaut et al. 2018}), and outreach to vulnerable populations (\cslcitation{42}{Wilde et al. 2021})---a cross-domain pattern that Amarasinghe et al. (\cslcitation{2}{Amarasinghe et al. 2025}) retrospectively synthesize as common methodological lessons across DSSG/DSaPP deployments. The Triage pipeline enforces temporal validation, evaluates at precision and recall at k, and integrates fairness auditing via Aequitas (\cslcitation{35}{Saleiro et al. 2018}). Rodolfa et al. (\cslcitation{34}{Rodolfa et al. 2020}; \cslcitation{33}{Rodolfa, Lamba, \& Ghani 2021}) demonstrated, first as a single FAT* case study and subsequently across multiple deployments, that fairness-accuracy tradeoffs in public policy ML are often negligible---a finding we explain through the multi-objective thesis: when the full space of data product configurations is explored, the Pareto front between fairness and performance is richer than it appears from any single model. Amarasinghe et al. (\cslcitation{3}{Amarasinghe et al. 2024}) showed that explainable ML methods fail to demonstrate utility in application-grounded contexts, directly supporting our performative evaluation thesis. Black et al. (\cslcitation{8}{Black et al. 2023}) argued for pipeline-aware fairness, anticipating our expanded hyperparameter space thesis. The present paper provides the philosophical framework that unifies these findings.
\subsection{Performative Prediction}
\label{sec:org4549ae6}

Perdomo et al. (\cslcitation{27}{Perdomo et al. 2020}) formalized performative prediction, showing that when predictions inform decisions, the act of prediction changes the distribution it aims to predict. They proposed \emph{performative stability}---a fixed point where predictions are calibrated against the outcomes resulting from acting on them. This formalization captures mathematically what our Thesis II describes ontologically. However, the performative prediction literature treats performativity as a technical challenge to be solved. Our framework reverses this: performativity is the fundamental condition of all applied ML, and the appropriate response is continuous institutional engagement, not equilibrium-seeking.
\subsection{Performativity in Economic Sociology}
\label{sec:org5758dd0}

Callon (\cslcitation{10}{Callon 1998}, \cslcitation{11}{2007}) argued that economic theories actively constitute markets. MacKenzie \& Millo (\cslcitation{24}{MacKenzie \& Millo 2003}) provided the canonical demonstration: the Black-Scholes-Merton model ``made itself true'' by shaping trader behavior. MacKenzie (\cslcitation{23}{MacKenzie 2006}) distinguished generic performativity (a model is used), effective performativity (a model changes behavior), and Barnesian performativity (a model makes the world resemble itself). We claim that deployed ML systems exhibit at least effective performativity.
\subsection{Cybernetic Ontology}
\label{sec:org5b38aad}

Pickering's (\cslcitation{28}{Pickering 2010}) account of British cybernetics distinguishes representational from performative ontologies. For Pickering, the cyberneticians (Ashby (\cslcitation{4}{Ashby 1956}), Beer (\cslcitation{7}{Beer 1972}), Pask (\cslcitation{26}{Pask 1976}), Bateson (\cslcitation{5}{Bateson 1972})) understood systems as fundamentally unknowable in the representational sense and proposed adaptive, experimental engagement. Beer's Viable System Model is particularly relevant to our Thesis XII: it insists on continuous feedback between the regulatory system and the system being regulated. Ashby's homeostat, which seeks viability rather than optimality, anticipates our Thesis VI on satisficing.
\subsection{Bounded Rationality and Satisficing}
\label{sec:org95d69bb}

Simon's (\cslcitation{38}{Simon 1956}, \cslcitation{39}{1972}) theory of bounded rationality argues that agents in complex environments cannot optimize; they satisfice---finding solutions that are good enough relative to their aspirations and constraints. This concept has been largely absent from the ML literature, which inherits an optimization-centric vocabulary from mathematical programming. Yet satisficing is the epistemologically appropriate stance for ML in complex adaptive systems: if the system coevolves with the model (Thesis II) and generalization is temporally bounded (Thesis V), then the optimal solution at time t may be suboptimal at t+1. What matters is not optimality but improvement over the current baseline---the organization's existing practice.
\subsection{Technical Debt and System-Level Concerns}
\label{sec:orgbf2f8ff}

Sculley et al. (\cslcitation{37}{Sculley et al. 2015}) identified hidden feedback loops, undeclared consumers, and data dependencies as sources of technical debt in ML systems. Our framework explains these at a deeper level: feedback loops are consequences of performativity (Thesis II), data dependencies reflect radical contextuality (Thesis III), and configuration issues arise from the expanded hyperparameter space (Thesis VIII). By providing a principled epistemological framework, we move beyond the metaphor of ``debt''---which implies a clean, debt-free state is achievable---toward understanding these as constitutive features of applied ML.
\section{The Thirteen Theses}
\label{sec:org9982707}

We present our framework as thirteen theses organized into four groups: epistemological (I, V, VI, X), ontological (II, III), methodological (VII, VIII, IX, XI), and political (IV, XII, XIII).\footnote{Definitions used throughout. \emph{Model} — the parameterized object produced by a learning procedure on training data. \emph{Data product} — the model together with its surrounding pipeline (feature stores, label-generation logic, deployment harness, monitoring infrastructure, retraining cadence, and the human operators and institutional structures that act on its output). \emph{Performative validity} — the criterion that a model is good if its deployment produces the effects its institution intends, evaluated against an explicit baseline of current practice (Theses IV, VI). \emph{Baseline} — the current organizational practice that the model is replacing or augmenting (a manual rule, a bureaucratic process, a status-quo allocation, or inaction), not a theoretical ideal or ``no-model'' abstraction. \emph{Materialist} — used here in the philosophical sense developed by Pickering (\cslcitation{28}{Pickering 2010}) in dialogue with the Marxian tradition: knowledge is constituted through situated practice and intervention, not detached observation, and the criterion of truth is the concrete effect of acting on the world.}
\subsection{Thesis I: The Model as Best Available Compression}
\label{sec:orgce97deb}

\emph{A well-trained model is the best representation of the system in question, given the available data.}

``Best representation'' does not mean ``true representation.'' Given the data available, the feature space the practitioner has chosen, and the algorithmic family imposed by the deployment context, the model is the best available compression of the system's \emph{exploitable regularities}. ``Best'' here is relative to these system constraints---not to truth, not to a hypothetical ideal model. This is less than what most practitioners assume (they often treat the model as revealing the system's structure) and more than what many critics concede (they sometimes dismiss ML models as mere curve-fitting). The same system, observed with a different feature set or modeled with a different algorithmic family, yields a different ``best'' compression---which is why feature engineering, cohort definition, and pipeline design are themselves epistemic acts (Thesis VIII), not mere preprocessing.
\subsection{Thesis II: The Data Product as Complex Adaptive System}
\label{sec:org962eb3c}

\emph{The data product---encompassing model training, deployment, monitoring, and retraining---is a complex adaptive system that coevolves with the system it models.}

\begin{figure}[H]
\centering
\begin{tikzpicture}[
  node distance=2.6cm,
  every node/.style={font=\small},
  box/.style={rectangle, draw, rounded corners=3pt, minimum width=2.6cm, minimum height=0.9cm, align=center, fill=blue!5, draw=blue!50!black},
  arr/.style={->, >={Stealth[length=2.4mm]}, thick, blue!60!black}
]
  \node[box] (model) {Model};
  \node[box, right=of model] (decision) {Decision};
  \node[box, below=of decision] (system) {Modified system};
  \node[box, below=of model] (data) {New data};

  \draw[arr] (model) -- node[above, font=\scriptsize, text=black] {informs} (decision);
  \draw[arr] (decision) -- node[right, font=\scriptsize, text=black] {modifies} (system);
  \draw[arr] (system) -- node[below, font=\scriptsize, text=black] {generates} (data);
  \draw[arr] (data) -- node[left, font=\scriptsize, text=black] {retrains} (model);
\end{tikzpicture}
\caption{The data product as a complex adaptive system. The model informs decisions; decisions modify the system being modeled; the modified system generates new data; the new data updates the model. The loop is continuous and recursive---there is no clean separation between observation and intervention.}
\label{fig:coevolution}
\end{figure}
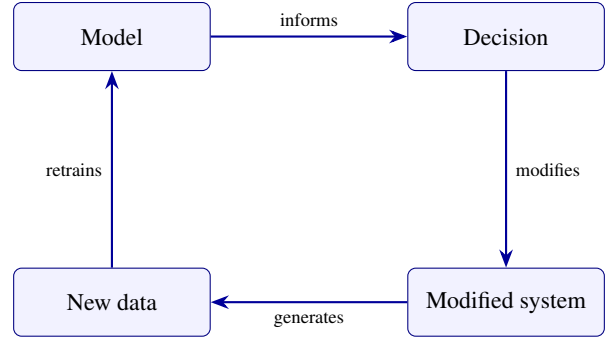

A model in production informs decisions, decisions modify the system, the modified system generates new data, and the new data modifies the model (Figure \ref{fig:coevolution}). This is coevolution in the sense of Holland (\cslcitation{19}{Holland 1995}). Two deployed public-sector systems illustrate the loop concretely: a Mexican social-services targeting system whose interventions reshape the household profiles it later sees (\cslcitation{36}{Sankaran et al. 2017}), and a clinical HIV-retention prediction system whose outreach decisions become part of the patient trajectories the next model is trained on (\cslcitation{22}{Kumar et al. 2020}). In both cases, the sequence \emph{outcome \(\to\) relabel \(\to\) retrain} is structural, not incidental. Concept drift is not a bug but the inevitable consequence of this coupling. Perdomo et al. (\cslcitation{27}{Perdomo et al. 2020}) formalize one aspect---how prediction shifts the data distribution. But our claim is broader: the entire data product, including its human operators, institutional context, and governance structures, constitutes a single complex adaptive system.

\subsection{Thesis III: Radical Contextuality}
\label{sec:orgc5296c2}

\emph{The model is radically contextual. It depends on the organization and the moment, and it is not transferable.}

If the model is a product of coevolution with a particular system, in a particular organization, at a particular moment, then transferring it is akin to transplanting an organ without considering the recipient's immunology. This contradicts the narrative of transferable solutions and the assumption that a model trained at organization A will perform at organization B. What may transfer are the most abstract structural features; the specific patterns of coupling between model, organization, and context do not.

An apparent counterexample is the success of foundation models---large pre-trained systems that demonstrably transfer across tasks and contexts. But this objection conflates two levels. What transfers in foundation models are low-level representations: visual features, syntactic patterns, distributional semantics. These capture regularities shared across contexts and correspond to what we acknowledge as ``the most abstract structural features.'' What does \emph{not} transfer is the specific coupling between model, organization, and decision context that constitutes the data product. A foundation model fine-tuned for one organization's operations will not work for another without re-engineering the entire pipeline: different data, different edge cases, different institutional definitions of success, different resource constraints. The foundation model provides initialization; the data product remains radically local. Moreover, foundation models themselves embed the political choices of their training data---which texts were included, which excluded, whose language is represented---confirming rather than refuting the materialist thesis.
\subsection{Thesis IV: Performative Validity}
\label{sec:org71fbc8e}

\emph{ML must be judged performatively: by its effects in the world, not by formal properties of the model.}

A model is good if its deployment produces the desired effects. \emph{Whose} desired effects, and against \emph{what} baseline, are themselves political questions: the choice of objective is the responsibility of stakeholders (Thesis XII), and validity must always be evaluated against an explicit baseline of current organizational practice (Thesis VI), not against a hypothetical optimum or a ``no-model'' counterfactual that cannot be observed once the model is deployed. This applies to prediction (did decisions improve?), explanation (did actors make better decisions because of it?), and ethics (did deployment reduce the injustices it targeted?). Amarasinghe et al. (\cslcitation{3}{Amarasinghe et al. 2024}) provide direct empirical support: explainable ML methods that appear beneficial under simplified evaluations fail in application-grounded contexts. Carton et al. (\cslcitation{12}{Carton et al. 2016}) illustrate the performative effect itself: early-intervention systems that score police officers for adverse-event risk are designed to reshape supervisor and officer behavior once deployed---the performative effect \emph{is} the design target, not a confound to be controlled for. The criterion is what the model \emph{does}, not what it \emph{is}.
\subsection{Thesis V: Operational, Not Scientific, Generalization}
\label{sec:org6e41efc}

\emph{Institutional support ML seeks operational generalization---continued performance in the population and time window of deployment---rather than scientific generalization, the discovery of stable structure in the world.}

These two goals require different validation regimes and different criteria of success. Operational generalization asks: will the model continue to work here, next month, on the cases I will actually see? Scientific generalization asks: does this finding replicate, in this form, at other times and places? Breiman (\cslcitation{9}{Breiman 2001}) began this argument by distinguishing algorithmic from data modeling cultures, but remained within a framework where predictive accuracy is the ultimate criterion---which is still the scientific frame. We push further: in the institutional applied context, even predictive accuracy is local and transient, and this is appropriate to a system whose purpose is intervention rather than discovery. Confusing the two regimes produces over-claiming (treating a model that works in one office as if it had discovered something universal) or, more rarely, under-claiming (refusing to deploy a model that demonstrably works locally because it lacks scientific generality).
\subsection{Thesis VI: Satisficing, Not Optimizing}
\label{sec:org3b13595}

\emph{The model does not need to be perfect. It needs to be better than the baseline---what the organization is actually doing now.}

The baseline is not a theoretical ideal. It is the organization's current practice: a manual rule, a bureaucratic process, an informal bias, or inaction. If the model outperforms that, it is useful. This is satisficing in Simon's (\cslcitation{38}{Simon 1956}) sense: the epistemologically appropriate decision strategy in complex environments where the optimal solution is either undefined, unstable, or computationally intractable. In a coevolutionary system (Thesis II) where generalization is temporally bounded (Thesis V), the ``optimum'' moves as you seek it. What matters is viability and improvement, not optimality.

Bauman et al. (\cslcitation{6}{Bauman et al. 2018}) make this concrete. Their pretrial-incarceration-reduction system was evaluated not against an idealized utility function but against the existing prioritization rule used by court services. The model ``succeeded'' by producing a smaller and more equitable holding population than the status-quo allocation---a satisficing goal that the organization could recognize and act on. An optimization-framed evaluation against a hypothetical optimum would have produced no actionable comparison; the value of the system lay precisely in being measurably better than the rule it replaced.

This has a political dimension. Defining the baseline forces the organization to make its current practice explicit---including the informal biases, ad-hoc rules, and institutional inertia that constitute the real status quo. Many organizations resist this because the baseline, once made visible, is often embarrassing. But without an explicit baseline, evaluation is impossible, and the demand for ``sufficiently good'' models in the abstract becomes an indefinitely movable goalpost that prevents deployment.
\subsection{Thesis VII: Temporality Is Inescapable}
\label{sec:org6f3042f}

\emph{Worthwhile ML problems involve an action to be taken and always occur in time. Cross-validation must always be temporal.}

If there is no concrete action that follows from the model's output, there is no ML problem worth solving. And if there is action, there is temporality. This yields a methodological imperative \emph{for any institutional applied context}: validation must be temporal. Shuffled cross-validation simulates a world where time does not exist, in order to evaluate a system that only exists in time. This is not merely a technical error---it is an epistemological contradiction. (Where the deployment context is genuinely atemporal---offline benchmarks against fixed test sets, scientific ML on stationary data---the imperative softens, but those are not the cases this paper addresses; see Thesis V.)
\subsection{Thesis VIII: The Expanded Hyperparameter Space}
\label{sec:org61bf59f}

\emph{The real hyperparameter space includes the entire data product: algorithm, hyperparameters, features, cohort definition, imputation strategy, data sources, temporal windows, and retraining frequency.}

Every decision in the data product's construction is a hyperparameter. Black et al. (\cslcitation{8}{Black et al. 2023}) recognize this in the fairness domain. We generalize: every property of interest---performance, fairness, robustness, temporal stability---is a property of the full pipeline, and the search through the expanded hyperparameter space is itself an epistemic process.
\subsection{Thesis IX: The Problem Is Almost Always Multi-Objective}
\label{sec:org359f340}

\emph{Almost any institutional ML problem is multi-objective in practice, driven first by finite resources and second by the plurality of values the system must serve. The algorithm optimizes a single scalar loss function; the data product navigates the full multi-objective space.}

The first reason a deployed problem is multi-objective is structural: \emph{resources for action are finite}. A precision@k threshold (Thesis XI) is itself an objective constraint, and the choice of k trades volume against quality. Once resources bind, no single scalar metric captures the deployment-relevant performance of the system. The second reason is \emph{value-pluralism}: even absent a hard resource cap, an institutionally deployed model must balance precision against equity across demographic groups, stability over time, operational interpretability, and implementation cost. No single algorithm optimizes all of these simultaneously. It is the data product---through the process of training multiple models, evaluating across multiple metrics, and selecting according to stakeholder priorities---that navigates the multi-objective space. (The genuine single-objective case---an offline benchmark with no resource constraint and a single homogeneous stakeholder---exists, but is rare outside of academic competitions and is not the setting this paper addresses.)

This has a direct empirical consequence. Rodolfa et al. (\cslcitation{34}{Rodolfa et al. 2020}; \cslcitation{32}{Rodolfa, Lamba, \& Ghani 2020}, \cslcitation{33}{2021}) demonstrated---first as a FAT* case study on misdemeanor recidivism, then as a broader cross-domain methodological analysis, and subsequently across multiple public-policy deployments---that fairness-accuracy tradeoffs are often negligible. Our framework explains why: when the full space of data product configurations is explored (Thesis VIII), the Pareto front between fairness and performance is far richer than it appears from any single model. The perceived inevitability of fairness-accuracy tradeoffs is an artifact of evaluating individual algorithms rather than the data product as a whole. This connects to the impossibility results of Chouldechova (\cslcitation{13}{Chouldechova 2017}) and Kleinberg et al. (\cslcitation{21}{Kleinberg, Mullainathan, \& Raghavan 2016}), which demonstrate that certain fairness criteria cannot be simultaneously satisfied---further evidence that the problem is irreducibly multi-objective and that navigating the tradeoff space is a political, not merely technical, task. This also further undermines the epistemic authority of feature importance (Thesis X): the algorithm's feature importance reflects its unidimensional optimization, not the multi-dimensional structure of the real problem.
\subsection{Thesis X: Models as Mediated Epistemic Instruments}
\label{sec:orgf8eb21f}

\emph{Training multiple models is a method of investigation, not just model selection. But what models reveal is their interaction with the system, not the system itself. Feature importance is an artifact of the algorithm, not a property of the system.}

When a random forest outperforms logistic regression, this tells us something about the system's structure (nonlinearities, interactions)---but always mediated by the algorithm's constraints. Feature importance is the most dangerous conflation: it tells us how a particular algorithm distributed its predictive capacity among variables at a specific moment. A different algorithm yields different rankings. Moreover, feature importance is a population-level statistic: it speaks to averages, not to the specific individual on whom an action will be taken. Using feature importance as a substitute for decision-making confuses an artifact of the measurement instrument with a property of the measured.
\subsection{Thesis XI: Precision and Recall at k}
\label{sec:org445bbce}

\emph{The metrics that matter are precision@k and recall@k, because resources for action are finite.}

Resources are always finite. The question is never ``how good is the model overall?'' but ``how good is it in the top k elements I can act on?'' Global accuracy, AUC, and average F1 are noise relative to this question. And the choice of k is a political and institutional decision (Thesis XII): how many resources to allocate, to which problem, at what frequency. A model is not good or bad in the abstract; it is good or bad for a given level of resources.
\subsection{Thesis XII: Political Responsibility Belongs to the Stakeholder}
\label{sec:orgc29960a}

\emph{The choice of objectives, models, actions, and ethical criteria is the responsibility of the stakeholder. These choices may change as the system evolves under the model's influence.}

Objective functions, fairness definitions, and derived actions are political decisions. The data scientist provides the space of possibilities and the consequences of each choice; the stakeholder chooses. Because the model acts on the system and the system changes (Thesis II), these decisions must be continuously revisited. This demands an ongoing relationship between the technical team and the governance structure, consistent with Beer's (\cslcitation{7}{Beer 1972}) VSM requirement for a continuous feedback channel between the regulatory system and the system being regulated.
\subsection{Thesis XIII: This Perspective Is Materialist}
\label{sec:orga6efac7}

\emph{The preceding theses rest on three materialist commitments: models are products of material conditions, the criterion of validity is practice, and the systems modeled are constituted by historically situated social relations.}

First: the model is a product of concrete material conditions---the data available, the infrastructure that exists, the social and economic relations that generate those data. Second: the criterion of truth is practice---not correspondence with abstract reality, but the concrete effect of intervention. Third: the system being modeled is constituted by historically situated social relations---objective functions encode political choices that reflect interests. This materialist framing connects to computational approaches to planning (\cslcitation{14}{Cockshott \& Cottrell 1993}), where the question of \emph{what to optimize for whom} is recognized as irreducibly political. The data scientists have interpreted the data in various ways; the point is to transform the systems.
\section{Discussion}
\label{sec:org6237c1c}

\subsection{Unifying Disparate Practices}
\label{sec:org76e5980}

A central contribution of this framework is revealing that well-known best practices are not independent rules of thumb but consequences of a single epistemology. Temporal cross-validation follows from Theses V and VII. Precision@k follows from Thesis XI and finite resources. Pipeline-aware fairness follows from Thesis VIII. The critique of feature importance follows from Theses X and IX (the algorithm's unidimensional optimization cannot capture the multi-objective structure). Satisficing over optimizing follows from Theses II, V, and VI (coevolution, temporal boundedness, and the primacy of baseline comparison). The requirement for continuous stakeholder engagement follows from Theses II and XII. When practitioners adopt these practices without understanding their principled basis, they risk applying them inconsistently or abandoning them under pressure.

A natural objection to this framework is that radical contextuality (Thesis III) and the rejection of scientific generalization (Thesis V) lead to relativism---an anything-goes position where no cumulative knowledge is possible. This objection misreads the framework. What we propose is better described as \emph{situated universalism}: the principles are universal (temporal validation, baseline comparison, multi-objective evaluation, stakeholder governance, performative assessment), but each instantiation is local. This is analogous to how evidence-based medicine operates: the principles of clinical trial design are universal, but each treatment must be validated in each population, and a drug approved for one condition must be separately tested for another. The framework also permits cumulative knowledge at the meta-level: we can learn across projects that certain types of systems exhibit characteristic patterns---for instance, that fairness-accuracy tradeoffs are generally negligible when the full pipeline space is explored (\cslcitation{33}{Rodolfa, Lamba, \& Ghani 2021}), or that post-hoc explanations fail to improve decisions in application-grounded contexts (\cslcitation{3}{Amarasinghe et al. 2024}). What we cannot do is transfer a specific trained model from one organizational context to another and expect it to work without re-engineering the data product. This is not relativism; it is epistemic honesty about the scope of ML's claims.
\subsection{Implications for ML Education}
\label{sec:org3e6ceed}

ML courses typically teach algorithms, loss functions, and evaluation metrics in a framework that is implicitly representational, atemporal, and unidimensional. Students learn cross-validation without learning why it must be temporal; feature importance without its limitations; fairness metrics without recognizing that the choice of metric is political; and optimization without learning when satisficing is more appropriate. Our framework suggests a pedagogy that begins with the problem (what action, against what baseline?), proceeds through the epistemology (what can models know? what are the objectives?), and only then addresses the techniques. Rodolfa et al. (\cslcitation{31}{Rodolfa et al. 2019}) and Rodolfa \& Ghani (\cslcitation{30}{Rodolfa \& Ghani 2021}) have argued for an experience-centered approach that embodies many of these principles.
\subsection{Implications for Governance}
\label{sec:orgd72bc50}

If the data product coevolves with the system (Thesis II), if the problem is multi-objective (Thesis IX), and if political decisions must be continuously revisited (Thesis XII), then governance cannot be a one-time certification. It must be an ongoing institutional process, analogous to Beer's (\cslcitation{7}{Beer 1972}) algedonic channel: a mechanism for signaling when performance has degraded or when the context has shifted enough that the original objectives are no longer appropriate. Current regulatory approaches focusing on pre-deployment audits capture only one moment in the life of a coevolving system.

A sophisticated objection arises here: if the model changes the system it models (Thesis II), then any evaluation is against a distribution that will shift once the model is deployed, creating an apparent infinite regress. How can we ever evaluate a system whose deployment invalidates the conditions under which it was evaluated? This objection is technically precise but practically resolvable. The answer lies in combining Perdomo et al.'s (\cslcitation{27}{Perdomo et al. 2020}) concept of performative stability---seeking fixed points where predictions are calibrated against the outcomes that result from acting on them---with Beer's (\cslcitation{7}{Beer 1972}) cybernetic governance: continuous monitoring, institutional feedback loops, and governance structures empowered to revise objectives when the system's evolution demands it. The evaluation is not against a fixed ground truth but against a moving target, and the governance structure must be designed to detect when the target has moved enough that the model's configuration requires revision. This is not regress; it is the normal condition of any regulatory system operating in a complex environment. Static evaluation is insufficient; the response is dynamic governance, not the abandonment of evaluation.
\subsection{Scope, Limitations, and Future Directions}
\label{sec:org278746e}

An important clarification of scope: the theses advanced here target ML for decision support in institutional settings---government, public health, criminal justice, logistics, resource allocation. In these domains, the model's output informs an action, the action modifies the system, and the temporal, performative, and political dimensions we have described are inescapable. However, ML is also used as a scientific instrument in domains where the data distribution is genuinely stationary and the model's output does not feed back into the system---protein structure prediction, astronomical classification, materials science. In these settings, shuffled cross-validation may be appropriate, performativity is minimal, and the epistemological framework of classical statistics or Breiman's algorithmic culture may suffice. We do not claim that our theses apply universally to all uses of ML; we claim that they apply necessarily to all uses where ML informs decisions that affect people and institutions.

Several theses deserve formal treatment. The coevolutionary dynamics of Thesis II could be modeled by extending performative prediction (\cslcitation{27}{Perdomo et al. 2020}; \cslcitation{16}{Hardt \& Mendler-Dünner 2025}) to capture institutional dynamics---including multi-agent and decision-dependent game settings (\cslcitation{25}{Narang et al. 2023}) and the asymmetric power between platforms and the populations on which they predict (\cslcitation{17}{Hardt, Jagadeesan, \& Mendler-Dünner 2022}). The expanded hyperparameter space of Thesis VIII and the multi-objective navigation of Thesis IX could be formalized using compositional frameworks from applied category theory. The satisficing criterion of Thesis VI requires formalization of baseline definition and improvement thresholds. The performative evaluation of Thesis IV requires empirical protocols that are only beginning to emerge (\cslcitation{3}{Amarasinghe et al. 2024}). Case studies demonstrating the theses in operation---drawn from criminal justice (\cslcitation{18}{Helsby et al. 2018}; \cslcitation{40}{Sobrino et al. 2026}), electoral analytics (the infrastructure of the 2012 Obama campaign being a public-record example (\cslcitation{20}{Issenberg 2012})), maritime logistics (\cslcitation{41}{Villalobos et al. 2026}), public health (\cslcitation{29}{Potash et al. 2015}), and government interventions for vulnerable populations (\cslcitation{42}{Wilde et al. 2021}; \cslcitation{15}{Gaut et al. 2018})---would provide the empirical grounding that a programmatic paper necessarily lacks and represent the most important direction for future work.
\section{Conclusion}
\label{sec:org4ad3384}

We have proposed a framework for understanding machine learning as a performative materialist practice: an intervention in complex adaptive systems, whose validity is measured by its effects in the world, whose success is judged against the existing baseline rather than an abstract optimum, whose products are radically contextual, whose problems are inherently multi-objective, and whose design embeds political choices. The thirteen theses unify a set of methodological prescriptions---temporal validation, evaluation at k, pipeline-aware fairness, satisficing over optimizing, the multi-objective nature of real problems, the critique of feature importance---under a coherent epistemology that draws on cybernetics, bounded rationality, economic sociology, and historical materialism. We hope this framework contributes to a more self-aware and more honest ML practice: one that acknowledges what models can and cannot know, that takes responsibility for what models do, and that recognizes the political nature of the choices that shape every data product.
\section*{Acknowledgements}
\label{sec:org5b5b245}
We thank the Data Science for Social Good and the Data Science and Public Policy (DSaPP) communities, particularly Rayid Ghani and collaborators at the University of Chicago Crime Lab and Carnegie Mellon University, for the empirical context that shaped this framework. We are grateful to colleagues at the Escuela de Gobierno y Transformación Pública del Tecnológico de Monterrey for early feedback on the theses, and to the prosecutorial-governance team---Patricia Villa, Stephany Cisneros, Cristian Paul Camacho Osnay, Armando García Neri, and Israel Hernández---whose deployed work crystallised many of the methodological claims advanced here.
\section{References}
\label{sec:org7ce5d51}

\begin{cslbibliography}{1}{0}
\cslbibitem{1}{Ackermann, Klaus, Joe Walsh, Adolfo De Unánue, Hareem Naveed, Andrea Navarrete Rivera, Sun-Joo Lee, Jason Bennett, et al. 2018. “Deploying Machine Learning Models for Public Policy: A Framework.” In \textit{Proceedings of the 24th ACM SIGKDD International Conference on Knowledge Discovery and Data Mining (KDD)}, 15–22.}

\cslbibitem{2}{Amarasinghe, Kasun, Rayid Ghani, Andy Lai, Liliana Millan, \& Kit T. Rodolfa. 2025. “Lessons Learned from Designing, Developing, and Deploying Machine Learning Systems for Social Good.” In \textit{AAAI Workshop on AI for Public Missions}.}

\cslbibitem{3}{Amarasinghe, Kasun, Kit T. Rodolfa, Sérgio Jesus, Valerie Chen, Vahan Balayan, Pedro Saleiro, Pedro Bizarro, Ameet Talwalkar, \& Rayid Ghani. 2024. “On the Importance of Application-Grounded Experimental Design for Evaluating Explainable ML Methods.” In \textit{Proceedings of the AAAI Conference on Artificial Intelligence}, 38:20921–29. 19.}

\cslbibitem{4}{Ashby, W. Ross. 1956. \textit{An Introduction to Cybernetics}. London: Chapman \& Hall.}

\cslbibitem{5}{Bateson, Gregory. 1972. \textit{Steps to an Ecology of Mind}. New York: Ballantine Books.}

\cslbibitem{6}{Bauman, Matthew J., Kate S. Boxer, Tzu-Yun Lin, Erika Salomon, Hareem Naveed, Lauren Haynes, Joe Walsh, et al. 2018. “Reducing Incarceration through Prioritized Interventions.” In \textit{Proceedings of the 1st ACM SIGCAS Conference on Computing and Sustainable Societies (COMPASS)}, 1–8.}

\cslbibitem{7}{Beer, Stafford. 1972. \textit{Brain of the Firm}. London: Allen Lane.}

\cslbibitem{8}{Black, Emily, Rakesh Naidu, Rayid Ghani, Kit Rodolfa, Daniel Ho, \& Hoda Heidari. 2023. “Toward Operationalizing Pipeline-Aware ML Fairness: A Research Agenda for Developing Practical Guidelines and Tools.” In \textit{Proceedings of the 3rd ACM Conference on Equity and Access in Algorithms, Mechanisms, and Optimization (EAAMO)}.}

\cslbibitem{9}{Breiman, Leo. 2001. “Statistical Modeling: The Two Cultures.” \textit{Statistical Science} 16 (3): 199–231.}

\cslbibitem{10}{Callon, Michel. 1998. “Introduction: The Embeddedness of Economic Markets in Economics.” In \textit{The Laws of the Markets}, edited by Michel Callon, 1–57. Oxford: Blackwell.}

\cslbibitem{11}{———. 2007. “What Does It Mean to Say That Economics Is Performative?” In \textit{Do Economists Make Markets? on the Performativity of Economics}, edited by Donald MacKenzie, Fabian Muniesa, \& Lucia Siu, 311–57. Princeton University Press.}

\cslbibitem{12}{Carton, Samuel, Jennifer Helsby, Kenneth Joseph, Ayesha Mahmud, Youngsoo Park, Joe Walsh, Crystal Cody, Estella Patterson, Lauren Haynes, \& Rayid Ghani. 2016. “Identifying Police Officers at Risk of Adverse Events.” In \textit{Proceedings of the 22nd ACM SIGKDD International Conference on Knowledge Discovery and Data Mining (KDD)}, 67–76.}

\cslbibitem{13}{Chouldechova, Alexandra. 2017. “Fair Prediction with Disparate Impact: A Study of Bias in Recidivism Prediction Instruments.” \textit{Big Data} 5 (2): 153–63.}

\cslbibitem{14}{Cockshott, W. Paul, \& Allin F. Cottrell. 1993. \textit{Towards a New Socialism}. Nottingham: Spokesman Books.}

\cslbibitem{15}{Gaut, Garren, Andrea Navarrete, Laila Wahedi, Paul van der Boor, Adolfo De Unánue, Jorge Díaz, Eduardo Clark, \& Rayid Ghani. 2018. “Improving Government Response to Citizen Requests Online.” In \textit{Proceedings of the 1st ACM SIGCAS Conference on Computing and Sustainable Societies (COMPASS)}.}

\cslbibitem{16}{Hardt, Moritz, \& Celestine Mendler-Dünner. 2025. “Performative Prediction: Past and Future.” \textit{Statistical Science} 40 (3): 417–36.}

\cslbibitem{17}{Hardt, Moritz, Meena Jagadeesan, \& Celestine Mendler-Dünner. 2022. “Performative Power.” In \textit{Advances in Neural Information Processing Systems (NeurIPS)}.}

\cslbibitem{18}{Helsby, Jennifer, Samuel Carton, Kenneth Joseph, Ayesha Mahmud, Youngsoo Park, Andrea Navarrete Rivera, Klaus Ackermann, et al. 2018. “Early Intervention Systems: Predicting Adverse Interactions between Police and the Public.” \textit{Criminal Justice Policy Review} 29 (2): 190–209.}

\cslbibitem{19}{Holland, John H. 1995. \textit{Hidden Order: How Adaptation Builds Complexity}. Addison-Wesley.}

\cslbibitem{20}{Issenberg, Sasha. 2012. \textit{The Victory Lab: The Secret Science of Winning Campaigns}. New York: Crown.}

\cslbibitem{21}{Kleinberg, Jon, Sendhil Mullainathan, \& Manish Raghavan. 2016. “Inherent Trade-Offs in the Fair Determination of Risk Scores.” \textit{arXiv preprint arXiv:1609.05807}.}

\cslbibitem{22}{Kumar, Avishek, Arthi Ramachandran, Adolfo De Unánue, Christina Sung, Joe Walsh, John Schneider, Jessica Ridgway, Stephanie Masiello Schuette, Jeff Lauritsen, \& Rayid Ghani. 2020. “A Machine Learning System for Retaining Patients in HIV Care.” \textit{arXiv preprint arXiv:2006.04944}.}

\cslbibitem{23}{MacKenzie, Donald. 2006. \textit{An Engine, Not a Camera: How Financial Models Shape Markets}. Cambridge, MA: MIT Press.}

\cslbibitem{24}{MacKenzie, Donald, \& Yuval Millo. 2003. “Constructing a Market, Performing Theory: The Historical Sociology of a Financial Derivatives Exchange.” \textit{American Journal of Sociology} 109 (1): 107–45.}

\cslbibitem{25}{Narang, Adhyyan, Evan Faulkner, Dmitriy Drusvyatskiy, Maryam Fazel, \& Lillian J. Ratliff. 2023. “Multiplayer Performative Prediction: Learning in Decision-Dependent Games.” \textit{Journal of Machine Learning Research} 24: 1–56.}

\cslbibitem{26}{Pask, Gordon. 1976. \textit{Conversation Theory: Applications in Education and Epistemology}. Elsevier.}

\cslbibitem{27}{Perdomo, Juan C., Tijana Zrnic, Celestine Mendler-Dünner, \& Moritz Hardt. 2020. “Performative Prediction.” In \textit{Proceedings of the 37th International Conference on Machine Learning (ICML)}, 119:7599–7609. Pmlr.}

\cslbibitem{28}{Pickering, Andrew. 2010. \textit{The Cybernetic Brain: Sketches of Another Future}. University of Chicago Press.}

\cslbibitem{29}{Potash, Eric, Joe Brew, Alexander Loewi, Subhabrata Majumdar, Andrew Reece, Joe Walsh, Eric Rozier, Emile Jarpe-Ratner, Alejandro Gabriel, \& Rayid Ghani. 2015. “Predictive Modeling for Public Health: Preventing Childhood Lead Poisoning.” In \textit{Proceedings of the 21st ACM SIGKDD International Conference on Knowledge Discovery and Data Mining (KDD)}, 2039–47.}

\cslbibitem{30}{Rodolfa, Kit T., \& Rayid Ghani. 2021. “Taking Our Medicine: Standardizing Data Science Education with Practice at the Core.” \textit{Harvard Data Science Review} 3 (1).}

\cslbibitem{31}{Rodolfa, Kit T., Adolfo De Unánue, Matthew Gee, \& Rayid Ghani. 2019. “An Experience-Centered Approach to Training Effective Data Scientists.” \textit{Big Data} 7 (4): 249–61.}

\cslbibitem{32}{Rodolfa, Kit T., Hemank Lamba, \& Rayid Ghani. 2020. “Machine Learning for Public Policy: Do We Need to Sacrifice Accuracy to Make Models Fair?” \textit{arXiv preprint arXiv:2012.02972}.}

\cslbibitem{33}{———. 2021. “Empirical Observation of Negligible Fairness–Accuracy Trade-Offs in Machine Learning for Public Policy.” \textit{Nature Machine Intelligence} 3: 896–904.}

\cslbibitem{34}{Rodolfa, Kit T., Erika Salomon, Lauren Haynes, Iván Higuera Mendieta, Jamie Larson, \& Rayid Ghani. 2020. “Case Study: Predictive Fairness to Reduce Misdemeanor Recidivism through Social Service Interventions.” In \textit{Proceedings of the 2020 Conference on Fairness, Accountability, and Transparency (FAT*)}, 142–53.}

\cslbibitem{35}{Saleiro, Pedro, Benedict Kuester, Loren Hinkson, Jesse London, Abby Stevens, Ari Anisfeld, Kit T. Rodolfa, \& Rayid Ghani. 2018. “Aequitas: A Bias and Fairness Audit Toolkit.” \textit{arXiv preprint arXiv:1811.05577}.}

\cslbibitem{36}{Sankaran, Kris, Diego Garcia-Olano, Mobin Javed, Maria Fernanda Alcala-Durand, Adolfo De Unánue, Paul van der Boor, Eric Potash, Roberto Sánchez Avalos, Luis Iñaki Alberro Encinas, \& Rayid Ghani. 2017. “Applying Machine Learning Methods to Enhance the Distribution of Social Services in Mexico.” \textit{arXiv preprint arXiv:1709.05551}.}

\cslbibitem{37}{Sculley, D., Gary Holt, Daniel Golovin, Eugene Davydov, Todd Phillips, Dietmar Ebner, Vinay Chaudhary, Michael Young, Jean-François Crespo, \& Dan Dennison. 2015. “Hidden Technical Debt in Machine Learning Systems.” In \textit{Advances in Neural Information Processing Systems}, 28:2503–11.}

\cslbibitem{38}{Simon, Herbert A. 1956. “Rational Choice and the Structure of the Environment.” \textit{Psychological Review} 63 (2): 129–38.}

\cslbibitem{39}{———. 1972. “Theories of Bounded Rationality.” In \textit{Decision and Organization}, edited by C. B. McGuire \& Roy Radner, 161–76. Amsterdam: North-Holland.}

\cslbibitem{40}{Sobrino, Fernanda, Adolfo De Unánue T., Edgar Hernández, Patricia Villa, Elena Villalobos, David Aké, Stephany Cisneros, Cristian Paul Camacho Osnay, Armando García Neri, \& Israel Hernández. 2026. “Designing AI for Prosecutorial Governance: Case Prioritization and Statutory Oversight in Mexico.” \textit{arXiv preprint arXiv:2601.00396}.}

\cslbibitem{41}{Villalobos, Elena, Adolfo De Unánue T., Fernanda Sobrino, David Aké, Stephany Cisneros, Jorge Lecona, \& Alejandra Matadamaz. 2026. “Toward Reducing Unproductive Container Moves: Predicting Service Requirements and Dwell Times.” \textit{arXiv preprint arXiv:2604.06251}.}

\cslbibitem{42}{Wilde, Harrison, Lucia L. Chen, Austin Nguyen, Zoe Kimpel, Joshua Sidgwick, Adolfo De Unánue, Davide Veronese, Bilal Mateen, Rayid Ghani, \& Sebastian Vollmer. 2021. “A Recommendation and Risk Classification System for Connecting Rough Sleepers to Essential Outreach Services.” \textit{Data \& Policy} 3: e2.}

\end{cslbibliography}
\end{document}